\documentclass[twocolumn,showpacs,prd]{revtex4}
\usepackage{dcolumn}
\usepackage{bm}
\usepackage[normalem]{ulem}
\begin{document}

\title{Weak Interaction Neutron Production Rates in Fully Ionized Plasmas}
\author{A. Widom and J. Swain}
\affiliation{Physics Department, Northeastern University, Boston MA 02115}
\author{Y.N. Srivastava}
\affiliation{Physics Department \& INFN, University of Perugia, Perugia IT}

\begin{abstract}
Employing the weak interaction reaction wherein a heavy electron is captured by a 
proton to produce a neutron and a neutrino, the neutron production rate for neutral 
hydrogen gases and for fully ionized plasmas is computed. Using the Coulomb atomic 
bound state wave functions of a neutral hydrogen gas, our production rate results are 
in agreement with recent estimates by Maiani {\it et al}. Using Coulomb scattering state 
wave functions for the fully ionized plasma, we find a substantially enhanced neutron 
production rate. The scattering wave function should replace the bound state wave 
function for estimates of the enhanced neutron production rate on water plasma drenched 
cathodes of chemical cells.
\end{abstract}


\maketitle

\section{Introduction \label{com}}

In years past we have been working on weak interaction inverse beta 
decay\cite{WL1:2006,WL2:2007,Cirillo:2012} 
including electromagnetic interactions with collective plasma modes of motion. 
Our considerations have been recently 
criticized\cite{Ciuchi:2012}. Our neutron production rate\cite{WL1:2006,WL2:2007} is 
a factor of $\sim 300$ larger than that of Maiani\cite{Ciuchi:2012}. Our purpose is to point 
out the source of this difference so that the physical principles may be resolved. 

In Sec.\ref{NPM}, the calculation of neutron production for a neutral plasma of 
Ciuchi {\it et al}\cite{Ciuchi:2012} is briefly reviewed. Since the surface plasmas 
of hot cathodes within which neutron production is observed\cite{Cirillo:2012} are fully 
ionized, the neutral atomic gas case is not relevant. The irrelevant two body wave 
function\cite{Ciuchi:2012} employed for the neutral gas case should be replaced by 
the two body Coulomb wave function relevant to the fully ionized plasma. 
This is the usual fully ionized plasma situation, for example, in the study of the weak 
interaction electron capture reactions  
\begin{eqnarray}
({\rm general})\ \ \ \ e^- + \ ^A_ZX \to  \ ^A_{(Z-1)}X +\nu_e\ ,
\nonumber \\ 
({\rm special\ case})\ \ \ \ e^-+p^+ \to n+\nu_e\ , 
\label{intro1}
\end{eqnarray}  
in solar\cite{Bahcall:1962} physics. Scattering Coulomb wave functions also enter 
laboratory high energy\cite{Bardin:1994} physics. 
The case of the fully ionized plasma is discussed in 
Sec.\ref{FIPM}. In Sec.\ref{rnpr} our previous neutron production 
estimates\cite{WL1:2006,WL2:2007,Cirillo:2012} are verified employing the scattering 
Coulomb wave function.

In the concluding Sec.\ref{conc} we briefly indicate how collective many body interactions 
may modify the situation.

\section{Neutral Gas of Atoms \label{NPM}}

For a gas of neutral objects which consist of a heavy electron bound to a proton, 
the Coulomb wave function in the zero total momentum frame 
\begin{equation}
\psi_{e^- p^+}({\bf r})=\frac{e^{-r/a}}{\sqrt{\pi a^3}}\ \ \ \ \ \ \ \ 
a=\frac{\hbar^2}{me^2},
\label{npm1}
\end{equation}
wherein \begin{math} {\bf r}={\bf r}_{e^-}-{\bf r}_{p^+}  \end{math} and 
\begin{math} m  \end{math} is the reduced mass of the heavy electron. 
With the lowest order Fermi cross section for a heavy electron to scatter 
from a proton producing a neutron and a neutrino, 
\begin{eqnarray}
\tilde{\nu }=v\sigma=\frac{c}{2\pi }\left(\frac{G_Fm^2}{\hbar c}\right)^2
(g_V^2+3g_A^2)\times
\nonumber \\ 
\left(\frac{\hbar }{mc}\right)^2(\gamma^2-\gamma_{Threshold}^2).
\label{npm2}
\end{eqnarray}
If \begin{math}  n \end{math} denotes the number of bound neutral objects 
per unit volume, then the transition rate per unit time per unit volume to produce 
neutrons from the decay of the neutral objects 
\begin{eqnarray}
\varpi_0((e^-p^+)\to n+\nu_e)=nv\sigma |\psi_{e^- p^+}(0)|^2\ ,
\nonumber \\ 
\varpi_0=\left(\frac{n}{\pi a^3}\right)v\sigma =
\left(\frac{n\tilde{\nu}}{\pi a^3}\right).
\label{npm3}
\end{eqnarray}
Up to this point we are in agreement with the comment of 
Ciuchi {\it et al}\cite{Ciuchi:2012}. Our disagreement involves the more 
physical regime wherein the plasma is fully ionized. The particles are 
charged and not neutral and the wave function Eq.(\ref{npm1}) chosen by 
Ciuchi {\it et al}\cite{Ciuchi:2012} is thereby incorrect. The correct wave 
function is written below.

\section{Fully Ionized Plasma Modes \label{FIPM}}

For a fully ionized plasma, the constituents of the plasma are the charged 
heavy electron and the proton. We seek the scattering state production of 
neutrons 
\begin{equation}
e^-+p^+\to n+\nu_e.
\label{fipm1}
\end{equation} 
The wave function factor \begin{math} |\psi (0)|^2  \end{math} needed to 
include Coulomb attraction into the scattering is changed from the neutral 
plasma value \begin{math} 1/(\pi a^3)  \end{math}. The positive energy 
\begin{math}E=mv^2/2=\hbar^2k^2/2m  \end{math} 
scattering Coulomb wave function\cite{S. Fluge:1970} must  replace 
Eq.(\ref{npm1}) ; i.e. in terms of the Gamma function 
\begin{math} \Gamma (z)  \end{math} and the confluent hypergeometric 
function \begin{math} _1F_1(\xi ;\zeta ;z)  \end{math} 
\begin{eqnarray}
\psi ({\bf r})=e^{i{\bf k\cdot r}}\Big[e^{\pi /(2ka)}
\Gamma \left(1-\frac{i}{ka}\right)\times 
\nonumber \\ 
_1F_1\left(\frac{i}{ka};1;\frac{kr-{\bf k\cdot r}}{ka}\right)\Big].
\label{fipm2}
\end{eqnarray} 
If \begin{math} r\to 0 \end{math}, then 
\begin{equation}
|\psi (0)|^2=\frac{(2\pi e^2/\hbar v)}{1-exp(-(2\pi e^2/\hbar v))}\ .
\label{fipm3}
\end{equation} 
The neutron production rate per unit time per unit volume is then 
\begin{eqnarray}
\varpi(e^-+p^+\to n+\nu_e)=n^2 v\sigma |\psi(0)|^2=
n^2 \tilde{\nu } |\psi(0)|^2\ ,
\nonumber \\ 
\varpi=\frac{2\pi \alpha cn^2\sigma  }{1-exp(-2\pi c \alpha/ v)}\ ,
\label{fipm4}
\end{eqnarray}
wherein \begin{math} \alpha =e^2/\hbar c \end{math}. 

\section{The Neutron Production Ratio \label{rnpr}}

The ratio \begin{math} \varpi /\varpi_0  \end{math} of the neutron 
production rates per unit time per unit volume can be deduced from 
Eqs.(\ref{npm3}) and (\ref{fipm4}). Thermal averaging 
at a temperature small on the scale of the heavy electron mass 
\begin{math} k_BT\ll mc^2  \end{math} yields\cite{Bahcall:1962} the transition rate per unit time 
per unit volume for producing neutrons 
\begin{eqnarray}
\eta=\frac{\varpi}{\varpi_0}=2\pi^2\alpha na^3\left<\frac{c}{v}\right>,
\nonumber \\ 
\eta \approx 2\pi^2\alpha na^3 \sqrt{\frac{2mc^2}{\pi k_BT}}\ ,
\label{rnpr1}
\end{eqnarray}
where \begin{math} n \end{math} is the number of electrons per unit 
volume.

Previously\cite{WL2:2007} estimated temperatures of hydride 
cathodes \begin{math} T\sim 5\times 10^3\ ^oK  \end{math} 
are in agreement with the observed hot color of their brightly light emitting 
surfaces\cite{Cirillo:2012}. The resulting neutron production as described by 
Eq.(\ref{rnpr1}) is given by \begin{math} \eta \sim 5\times 10^2  \end{math}
in rough agreement with our previous estimates\cite{WL1:2006,WL2:2007, Cirillo:2012}. The 
factor of \begin{math} \sim 300 \end{math} discrepancy is thereby resolved.

\section{Conclusion \label{conc}}

Many body plasma effects on neutron production may be described by the correlations 
between the electron coordinates \begin{math} ({\bf r}_1, \cdots ,{\bf r}_N) \end{math}
and proton coordinates \begin{math} ({\bf s}_1, \cdots ,{\bf s}_N) \end{math} as 
given by the correlation function 
\begin{equation}
C=\frac{1}{N}
\left<\sum_{i=1}^N\sum_{j=1}^N \delta\big({\bf r}_i-{\bf s}_j\big)\right>=n\xi
\label{conc1}
\end{equation}
wherein \begin{math} \xi=|\psi (0)|^2 \end{math} only if there are merely two body 
collisions in the plasma. Collective oscillations and many body collisions would tend to 
raise the value of \begin{math} \xi \end{math} but require a many body Greens function 
analysis to include such effects in detail. However, previous discrepancies 
are now understandable.

We reiterate that at the level of dilute plasma two-body  correlations dealt with in
previous work\cite{Ciuchi:2012}, the order of magnitude of the discrepancy has herein 
been resolved.

\end{document}